\documentclass[a4paper,11pt]{article}
\pdfoutput=1
\usepackage{pos,bbm}
\usepackage{subfigure}
\newcommand{\Nff}{N_f^{(F)}}
\newcommand{\Nfa}{N_f^{(A)}}
\newcommand{\kf}{\ensuremath{\kappa_F}}

\title{Mixed adjoint-fundamental matter	and applications towards SQCD and beyond}

\author*[a,b]{Georg Bergner}
\author[c]{Stefano Piemonte}

\affiliation[a]{University of Jena, Institute for Theoretical Physics,\\ 
	Max-Wien-Platz 1, D-07743 Jena, Germany}
\affiliation[b]{University of M\"unster, Institute for Theoretical Physics,\\ 
	Wilhelm-Klemm-Str.~9, D-48149 M\"unster, Germany}
\affiliation[c]{University of Regensburg, Institute for Theoretical Physics,\\ 
	Universit\"atsstr.~31, D-93040 Regensburg, Germany}

\emailAdd{georg.bergner@uni-jena.de}

\abstract{Gauge theories with fermions in adjoint and fundamental representations are relevant for many different applications including composite Higgs models and general aspects of the confinement problem. We present first results from simulations of SU(2) gauge theory with two Dirac fermions in the fundamental representation and one adjoint flavor. In this context, we also discuss applications towards simulations of supersymmetric QCD.}

\FullConference{%
 The 38th International Symposium on Lattice Field Theory, LATTICE2021
  26th-30th July, 2021
  Zoom/Gather@Massachusetts Institute of Technology
}

\begin{document}
\maketitle

\section{Applications for gauge theories with adjoint-fundamental matter}
The simulations of gauge theories with fermions in two different representations has been subject of some recent studies. The main focus has been a composite Higgs model with SU(4) gauge theory coupled to fundamental and sextet fermions  \cite{DeGrand:2016pgq,Ayyar:2017qdf,Ayyar:2018zuk,Cossu:2019hse}.
Our simulations of a gauge theory with fermionic matter in adjoint and fundamental representation has three different basic motivations. 

The first goal are lattice simulation of supersymmetric gauge theories, which require adjoint fermions in the gauge sector. The second motivation is an investigation of composite Higgs scenarios, in which a near conformal theory is coupled to the Standard Model. A third target of our studies is the investigation of confinement in an enlarged space of theories, which includes QCD-like as well as supersymmetric gauge theories. This investigation provides new approaches for an analytic understanding of non-perturbative properties. These different motivations will be explained in the following.   

The basic lattice action consists of the standard Wilson plaquette gauge action $S_G$ coupled to $\Nff$ fundamental and $\Nfa$ adjoint Dirac fields,
\begin{align*}
\mathcal{S}_L =
S_G+\sum_{x,y}\sum_{n_f=1}^{\Nff} \bar{\psi}_x^{n_f}(D^{(F)}_w)_{xy}\psi_y^{n_f}+\sum_{x,y}\sum_{n_f=1}^{\Nfa} \bar{\psi}_x^{n_f}(D^{(A)}_w)_{xy}\psi_y^{n_f}\, ,
\end{align*}
where the Dirac operators  $D^{(F)}_w$ and $D^{(A)}_w$ are Dirac-Wilson operators with links in the fundamental and adjoint representation. We have added a clover term with a coefficient determined at one-loop approximation given in \cite{Musberg:2013foa}. We have seen in our most recent investigations of $\mathcal{N}=1$ supersymmetric Yang-Mills theory (SYM)  summarized in \cite{Ali:2019agk} that this one-loop approximation already provides a reasonable improvement. Note that half integer $\Nfa$ denote $2\Nfa$ Majorana fermions. 

This theory is closely related to supersymmetric QCD. In general, any supersymmetric gauge theory needs adjoint fermions as a supersymmetric partner for the gauge bosons. The minimal non-abelian gauge theory is $\mathcal{N}=1$ SYM with gluon fields and $\Nfa=1/2$ fermions, which means a Majorana gluino field. This pure gauge part can be extended by a matter sector. Each flavor needs fundamental fermion (quark) and scalar (squark) fields. The scalar fields drastically enlarge the space of tuning parameters, which means a larger number of observables need to be monitored in order to characterize the theory. Some attempts towards a simulations of SQCD are presented in 		\cite{Giedt:2009yd,Costa:2017rht,Wellegehausen:2018opt,Bergner:2018znw}.  The first simplification is hence to send the scalar mass to infinity and consider a theory with $\Nfa=1/2$  and $\Nff$ fermion fields. In the same way,  $\Nfa=1$ corresponds to $\mathcal{N}=2$ SYM with $\Nff$ flavors without scalar fields.

In composite Higgs scenarios, the Higgs sector of the Standard Model is generated by an additional strong interactions. There are severe restriction from experimental constraints for such strong interactions and a large number of theories have been investigated to find possible candidates. One notable example is SU(2) gauge theory with $\Nff=2$ fundamental Dirac fermions and results for this theory have been presented for example in \cite{Arthur:2016dir} and also at this conference \cite{DrachLattice21}. The theoretical considerations favor a scenario in which the theory is close the a conformal fixed point, which means the gauge coupling is nearly constant for a large energy range. Assuming a large mass anomalous dimension is this energy range, it is possible to accommodate mass generation and electroweak symmetry breaking. Compared to the fundamental representation, a near conformal behavior can be reached with a lower number of adjoint fermion flavors. Therefore especially the $\Nfa=2$ theory has been considered as an interesting candidate. 
However, the mass anomalous dimension is not large in this case. 
More recently $\Nfa=1$ has been found to be at least near conformal with a significantly larger mass anomalous dimension.
In this case, the field content is, however, not sufficient to break the electroweak symmetry of the Standard Model. A further discussion of these adjoint theories can be found in \cite{Athenodorou:2014eua} and has also been presented at this conference \cite{Bennett:2021ivn}.

The theory considered here with $\Nff=2$ fundamental and $\Nfa=1$ Dirac flavors is an ideal extension of these studies. It is closely related to the pure fundamental $\Nff=2$ case and can be seen as a modification of $\Nfa=1$ towards a realistic extension of the Standard Model. It has been proposed much earlier in \cite{Ryttov:2008xe} as a possible minimal way to extend the Standard Model.

There is a third and completely different motivation for an investigation of SU(2) gauge theory with mixed adjoint-fundamental matter in terms of the general aim to provide an understanding of the confinement mechanism. 
Based on supersymmetry and semiclassical approaches, there is a good understanding of the mechanism for compactified gauge theories with adjoint fermions. In $\mathcal{N}=1$ SYM, the Witten index remains constant when going from the compactified to the full theory. It has been conjectured that there is a more general continuity between confinement in Yang-Mills theory and the compactified regime with adjoint fermions. Towards an understanding of full QCD, the addition of fundamental matter is required. Especially SU(2) gauge theory with $\Nfa=1/2$ and $\Nff=2$ fermions is an interesting candidate. It provides a color-flavor center symmetry which is preserved even with the fundamental fermions. This means deconfinement remains a real phase transition and does not become a cross-over. A more complete discussion of the theoretical background can be found in \cite{Kanazawa:2019tnf} and lattice studies of the compactified SYM are presented in \cite{Bergner:2018unx,Bergner:2014dua}.

These different approaches to the Standard Model and to the confinement problem provide a reasonable motivation for an investigation of theories with adjoint and fundamental matter. Our current studies are only the first step in this direction corresponding to a scan of the basic parameters and consistency checks. Results have been presented in \cite{Bergner:2020mwl}. We are also interested to find out how close the theory is to a conformal behavior. This is very important for any more detailed investigation since a much different dependence on the masses compared to a QCD-like behavior is expected for the conformal case.

\section{Numerical lattice simulations}
We have done numerical simulations of the theory with one-loop clover improved Wilson fermions. The clover coefficients are determined by the expressions derived in Ref.~\cite{Musberg:2013foa} for fermions in fundamental and adjoint representation. The Wilson fermions require a tuning of the fundamental ($m_0$) and adjoint $(m_0)_A$ mass parameters. One important first task is to determine the dependence of renormalized masses on the bare parameters. Two independent PCAC relations can be defined for the two representations, which leads to $m_{PCAC,F}$ for the fundamental and $m_{PCAC,A}$ for the adjoint representation. If the theory is QCD-like, a chiral symmetry breaking breaking pattern of the following form is expected due to fundamental and adjoint fermion condensates,
\begin{align*}
	\text{SU}(2\Nff)\rightarrow \text{Sp}(2\Nff)\, ,\quad 	\text{SU}(2\Nfa)\rightarrow \text{SO}(2\Nfa)\, .
\end{align*}
In addition, a near conformal scaling might also describe the low energy effective theory.

\begin{figure}
	\subfigure[Pure adjoint theory\label{bulk_adj}]
	{\includegraphics[width=0.47\textwidth]{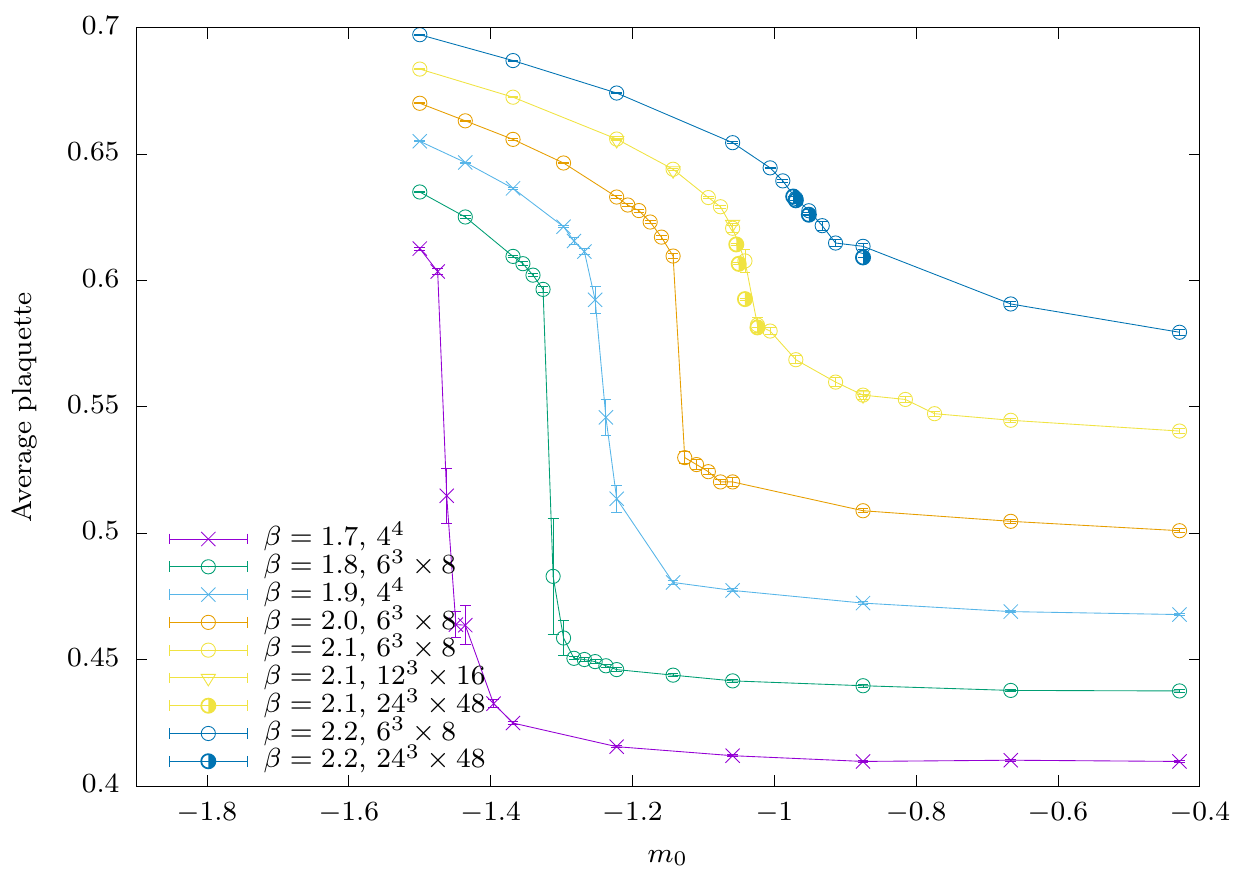}}
	\subfigure[$\kf$ dependence\label{bulk_kf}]
	{\includegraphics[width=0.47\textwidth]{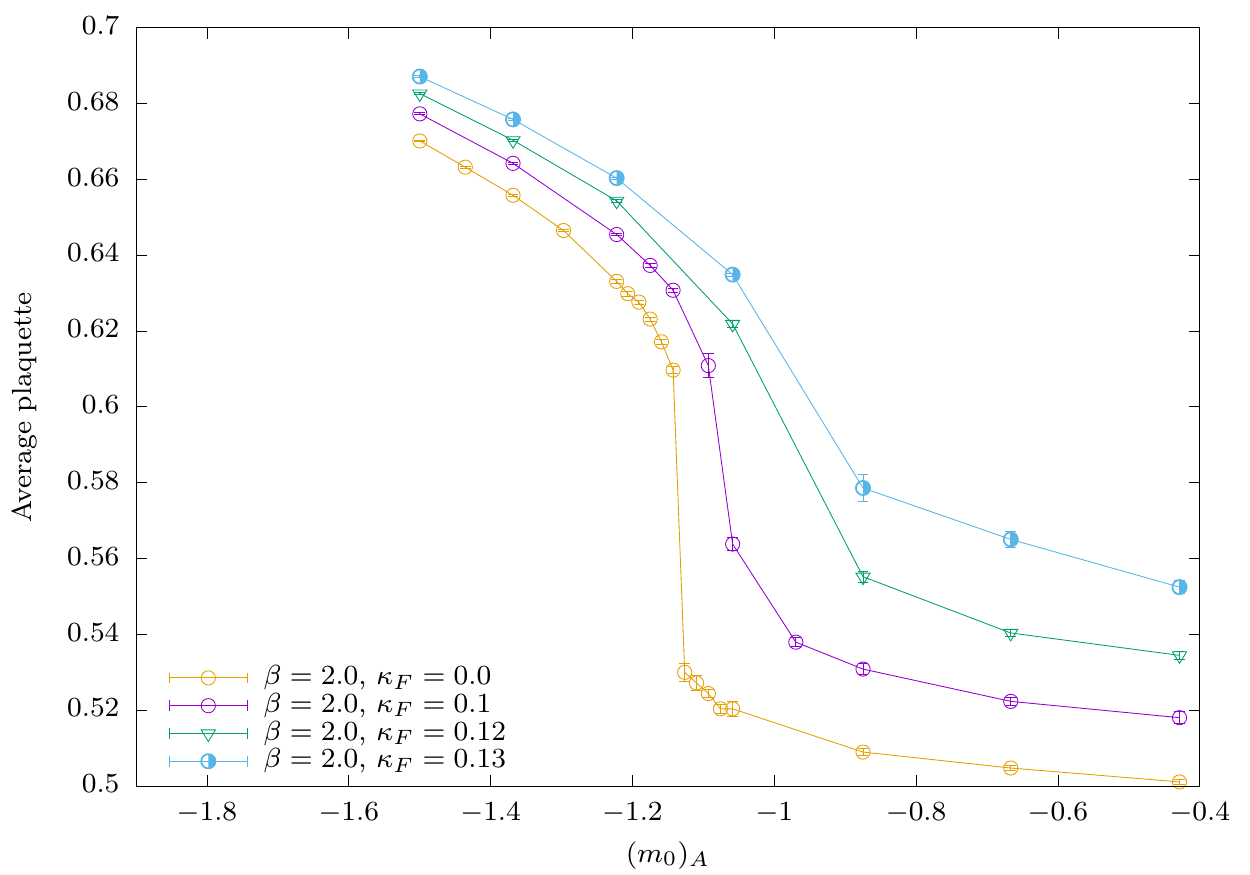}}
	\caption{Bulk phase transitions indicated by discontinuity of the average plaquette as a function of the bare mass parameter. Figure \ref{bulk_adj} shows the $\Nff=0$ case and the main simulations have been done on a $N_s\times N_t=4^4$ lattice. The dependence on mass of with additional fundamental fermions is shown in Figure \ref{bulk_kf}. \label{fig:bulk} }
\end{figure}
Before the main numerical simulations can be started, it is necessary to investigate the general parameter space to exclude possible bulk phases. These phases are induced by lattice artifacts and have been observed for several theories with adjoint fermions. As shown in Fig.~\ref{fig:bulk}, the transition can be observed as a discontinuity of the average plaquette value. The additional fundamental fermions reduce the strength of the transition, but, to be on the safe side, we limit our gauge coupling to the range $\beta\geq 2.1$.

\begin{figure}
	\centerline{\includegraphics[width=0.5\textwidth]{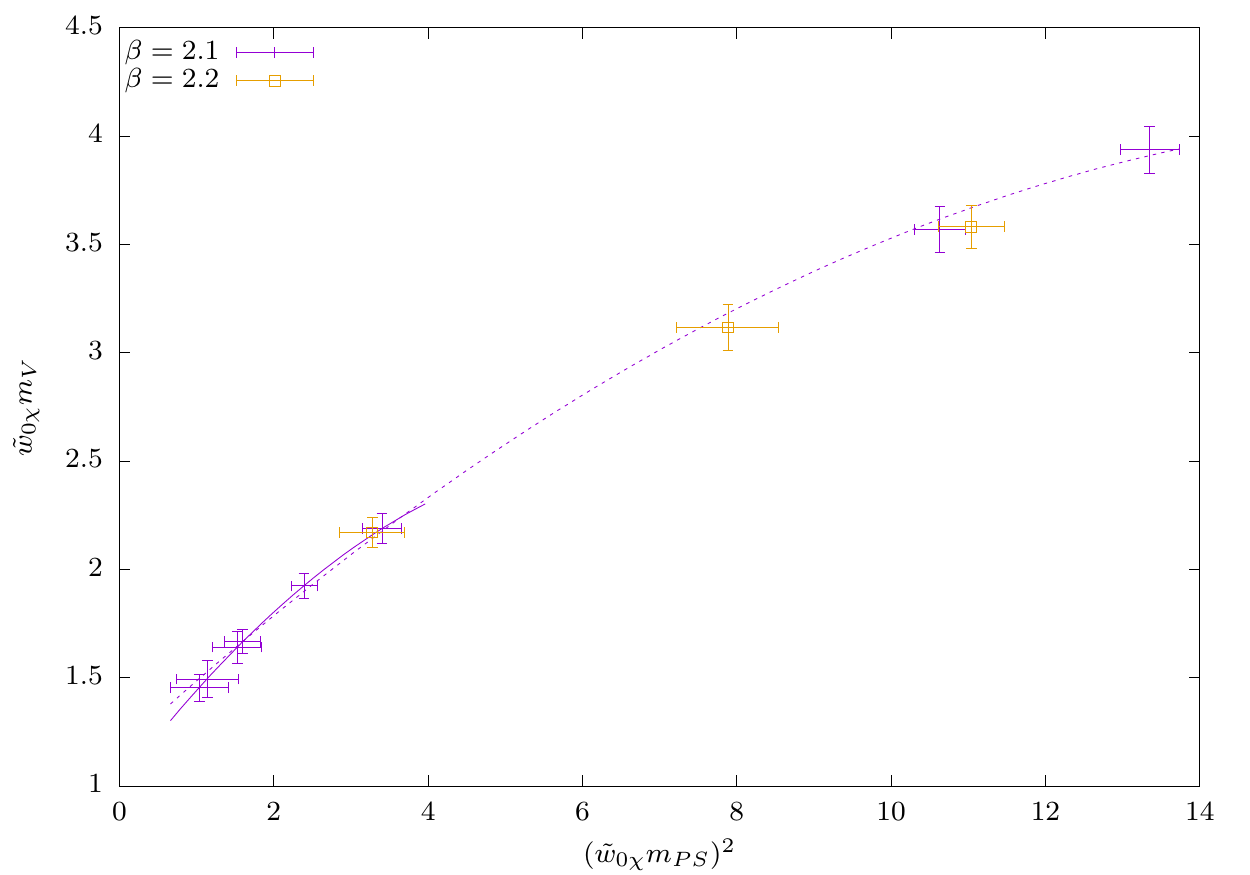}}
	\caption{The vector meson mass of the SU(2) $\Nff=2$ fundamental theory extrapolated to the chiral limit. A quadratic fit function is used for the $\beta=2.1$ data. The dotted line indicates the fit of the complete range, but the final value is obtained only in the range $(\tilde{w}_{0\chi} m_{PS})^2<4$ indicated by the solid line. The $\beta=2.2$ data is added for comparison.  \label{fig:vec} }
\end{figure}
In addition, we have performed comparisons with other available data as consistency checks. Precise data are available in the pure fundamental limit, which means SU(2) gauge theory with $\Nff=2$ fundamental fermions. We have compared our chiral extrapolation of the vector mass in units of the gradient flow scale $\tilde{w}_{0\chi}$, see Fig.~\ref{fig:vec}. Our result $\tilde{w}_{0\chi} m_{V\chi}=1.008(9)$ is compatible with the known continuum extrapolation $\tilde{w}_{0\chi} m_{V\chi}=1.01(3)$ of \cite{Arthur:2016dir}.

\subsection{Chiral symmetry breaking}
\begin{figure}
	\subfigure[Fundamental representation\label{mpsf}]
	{\includegraphics[width=0.47\textwidth]{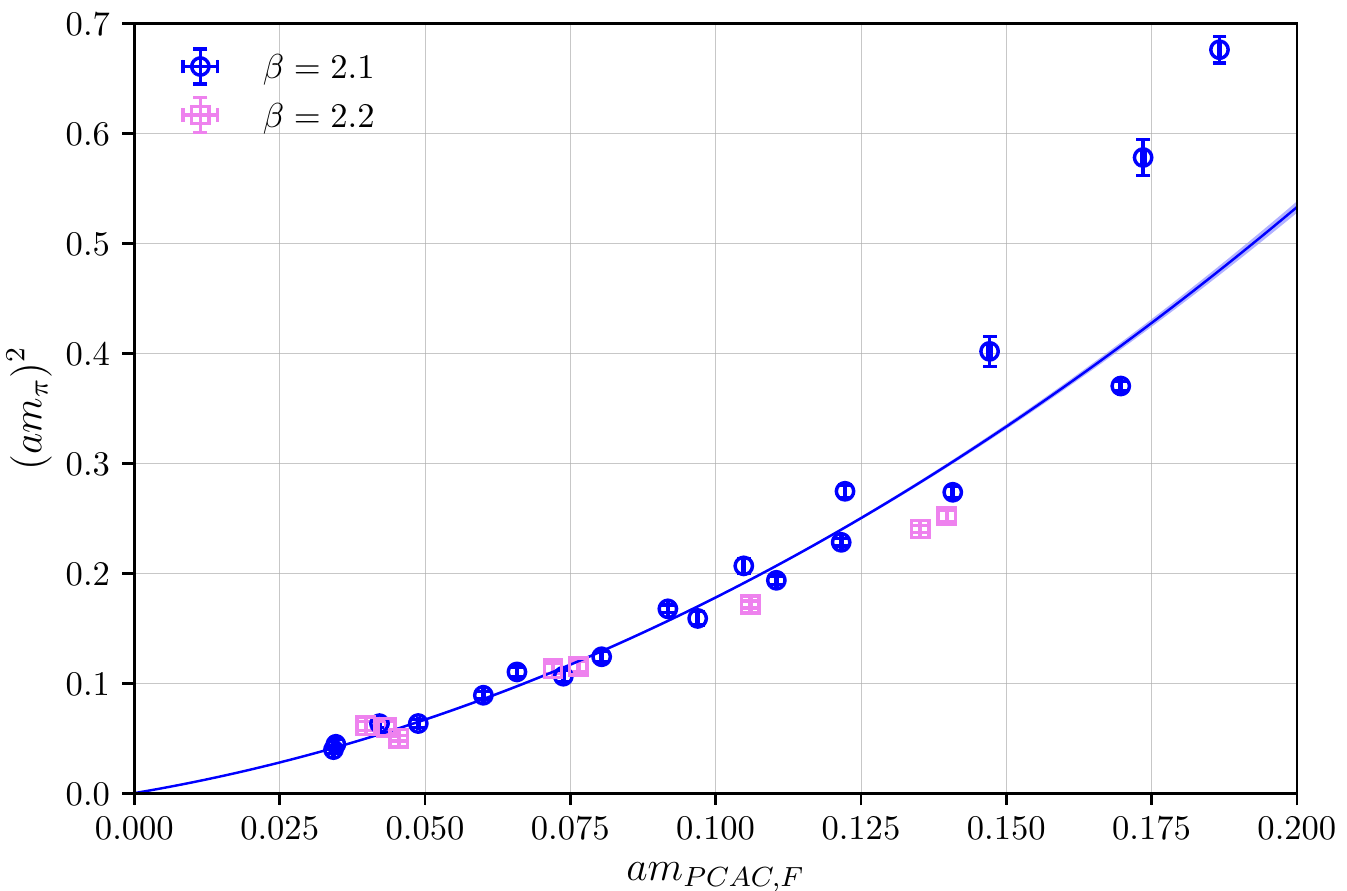}}
	\subfigure[Adjoint representation\label{mpsa}]
	{\includegraphics[width=0.47\textwidth]{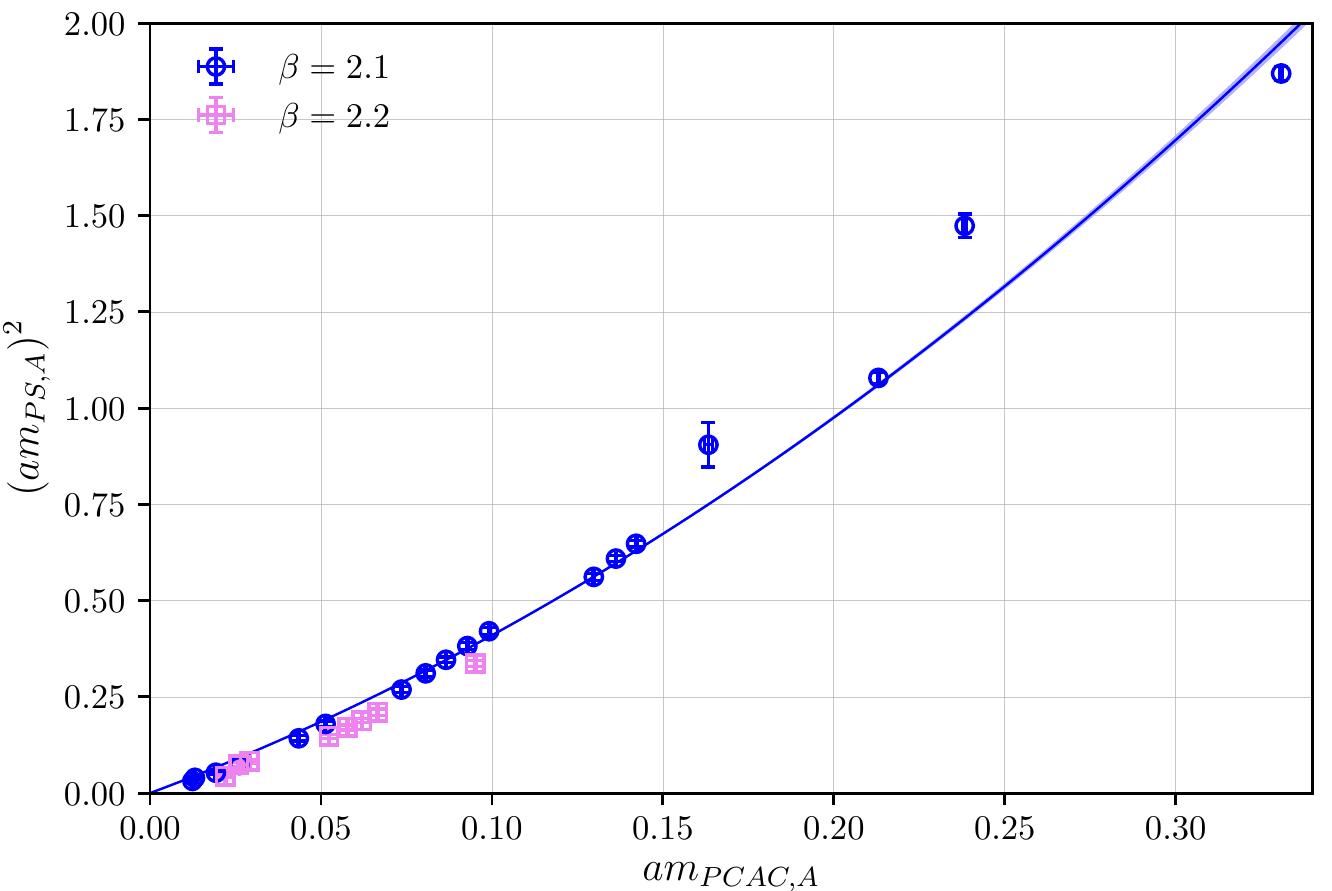}}
	\caption{The dependence of the pseudoscalar meson mass in fundamental ($m_\pi$) and adjoint ($m_{PS,A}$) mass on the PCAC mass in the same representation.\label{fig:chiral_fit}}
\end{figure}
The simplest scenario for the low energy effective theory is a QCD-like chiral symmetry breaking, meaning that bound states scales with the pseudo Nambu-Goldstone (pion) mass according to a generalized chiral perturbation theory.
In the first simplified approach, one can assume that the main dependence of the meson masses and decay constants is given by the mass in the same representation. As shown in Fig.~\ref{fig:chiral_fit}, our data shows an approximate agreement with this assumption.

The simplified fit can be extended to include the relevant contribution from the other representation. This dependence  has been worked out in \cite{DeGrand:2016pgq}. In our study, we can only focus on few dominant contributions since the determination of the large number of fit parameter requires more data. However, the fit of these main parameters agrees reasonably well with our data.

\subsection{Conformal scaling}
The second infrared scenario is a conformal scaling. In this case, the masses of all hadrons are expected to vanish if the two renormalized masses are sent to zero. The scaling of hadron correlation function at a scale change by $\mu\rightarrow \mu/b$ ($b>1$) is
\begin{align*}
C_H(t,g_i,m_i,\mu)=b^{-2y_H}C_H(t/b,b^{y_{g_i}}g_i,b^{y_i}m_i,\mu/b)
\end{align*}
with scaling exponents $y_{g_i}$, $y_i$, and $y_H$. In the conformal case, the relevant directions are given by the mass parameters with scaling dimensions $y_i$. If there is only a single mass parameter and assuming a $\exp(-M_Ht)$ dependence of the correlator, one obtains the usual scaling of hadron masses as $M_H\sim m^{1/y}$.

\begin{figure}
	\centerline{\includegraphics[width=0.7\textwidth]{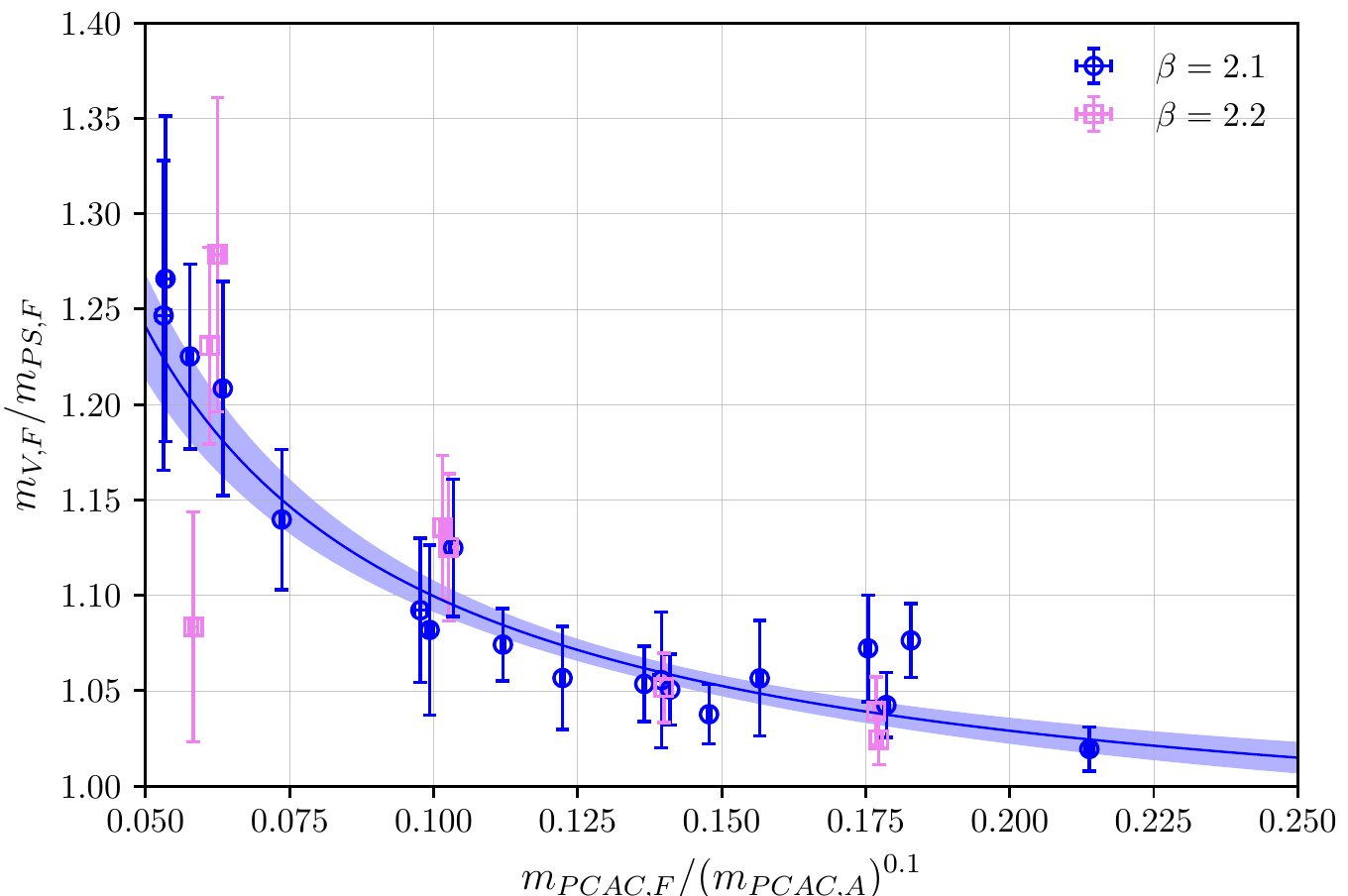}}
	\caption{The ratio of vector over pseudoscalar mass in the fundamental representation as a function of $am_{PCAC,F} (am_{PCAC,A})^{-r}$. The line shows a fit according to $a+b/x$ of the $\beta=2.1$ data. The value of $r=0.1$ corresponds to an optimal choice according to the residual. \label{fig:confrat}}
\end{figure}
If fermions are in the same representation like in \cite{Hasenfratz:2016gut}, masses have the same scaling dimension, which simplifies the expressions.
In our case, we have to consider two different scaling dimensions $y_F$ and $y_A$ for adjoint and fundamental representations, which leads to
\begin{align*}
\frac{am_{V,F}}{am_{PS,F}}=F_R(am_{PCAC,F} (am_{PCAC,A})^{-y_F/y_A})\, ,
\end{align*}
for ration of pseudoscalar $am_{V,F}$ and vector meson $am_{PS,F}$ in the fundamental representation with a general function $F_R$. Using a simple form for $F_R$, we have tried to fit the mass ratio, see Fig.~\ref{fig:confrat}. The best fit prefers $y_F/y_A=0.1$. In addition, we have determined the values of $y_F$ and $y_A$ by a fit of the dominant dependence of the hadron mass on the PCAC mass for the same representation ($M_H\sim m^{1/y}$). Since this dominant dependence leads to a much larger $y_F/y_A=0.1$, we conclude that there is not a completely consistent picture of conformal scaling. Nevertheless the mass ratios in Fig.~\ref{fig:confrat} are nearly constant as expected for a near conformal theory.

\section{Conclusions}
We have presented here the results of our simulations of SU(2) gauge theory with $\Nff=2$ fundamental and $\Nfa=1$ adjoint Dirac fermions. This theory represents a non-trivial extension of gauge theories considered so far and provides a number of interesting applications. It is related to supersymmetric gauge theories, composite Higgs models, and theories with analytically tractable confinement.

Due to the extended parameters space, the simulations are more demanding than in the simple case with only a single representation. We have performed initial tests and first numerical investigations of the theory. We observed indications rather towards a chiral symmetry breaking and not conformal behavior. The theory seems to be still quite close to the conformal case. Further investigations including a more detailed study of the particle spectrum are required to have a complete picture of the theory. 

\section*{Acknowledgements}

The authors gratefully acknowledge the Gauss Centre for Supercomputing e.V. (www.gauss-centre.eu) for funding this project by providing computing time on the GCS Supercomputer SuperMUC at Leibniz Supercomputing Centre (www.lrz.de).
Further computing time has been provided on the compute cluster PALMA of the
University of M\"unster. GB acknowledges support from the Deutsche Forschungsgemeinschaft (DFG)
under Grant No.\ BE 5942/3-1.



\begin{thebibliography}{99}
		\bibitem{DeGrand:2016pgq} 
		T.~DeGrand, M.~Golterman, E.~T.~Neil and Y.~Shamir,
		Phys.\ Rev.\ D {\bf 94}, no. 2, 025020 (2016)
		[arXiv:1605.07738 [hep-ph]].
		
		
		
		\bibitem{Ayyar:2017qdf} 
		V.~Ayyar, T.~DeGrand, M.~Golterman, D.~C.~Hackett, W.~I.~Jay, E.~T.~Neil, Y.~Shamir and B.~Svetitsky,
		Phys.\ Rev.\ D {\bf 97}, no. 7, 074505 (2018)
		[arXiv:1710.00806 [hep-lat]].
		
		
		
		\bibitem{Cossu:2019hse} 
		G.~Cossu, L.~Del Debbio, M.~Panero and D.~Preti,
		Eur.\ Phys.\ J.\ C {\bf 79}, no. 8, 638 (2019)
		[arXiv:1904.08885 [hep-lat]].
		
		\bibitem{Ayyar:2018zuk}
		V.~Ayyar, T.~Degrand, D.~C.~Hackett, W.~I.~Jay, E.~T.~Neil, Y.~Shamir and B.~Svetitsky,
		Phys. Rev. D \textbf{97} (2018) no.11, 114505
		[arXiv:1801.05809 [hep-ph]].
		
		\bibitem{Musberg:2013foa} 
		S.~Musberg, G.~Münster and S.~Piemonte,
		JHEP {\bf 1305}, 143 (2013)
		[arXiv:1304.5741 [hep-lat]].
		
		
		\bibitem{Ali:2019agk}
		S.~Ali, G.~Bergner, H.~Gerber, I.~Montvay, G.~M\"unster, S.~Piemonte and P.~Scior,
		Phys. Rev. Lett. \textbf{122} (2019) no.22, 221601
		doi:10.1103/PhysRevLett.122.221601
		[arXiv:1902.11127 [hep-lat]].
		
			
		
		
		\bibitem{Giedt:2009yd} 
		J.~Giedt,
		Int.\ J.\ Mod.\ Phys.\ A {\bf 24}, 4045 (2009)
		[arXiv:0903.2443 [hep-lat]].
		
		
		
		\bibitem{Costa:2017rht} 
		M.~Costa and H.~Panagopoulos,
		Phys.\ Rev.\ D {\bf 96}, no. 3, 034507 (2017)
		[arXiv:1706.05222 [hep-lat]].
		
		
		
		
		
		\bibitem{Wellegehausen:2018opt} 
		B.~Wellegehausen and A.~Wipf,
		PoS LATTICE {\bf 2018}, 210 (2018)
		[arXiv:1811.01784 [hep-lat]].
		
		

		\bibitem{Bergner:2018znw} 
		G.~Bergner and S.~Piemonte,
		PoS LATTICE {\bf 2018}, 209 (2019)
		[arXiv:1811.01797 [hep-lat]].
		
		
		
		\bibitem{Arthur:2016dir} 
		R.~Arthur, V.~Drach, M.~Hansen, A.~Hietanen, C.~Pica and F.~Sannino,
		Phys.\ Rev.\ D {\bf 94}, no. 9, 094507 (2016)
		[arXiv:1602.06559 [hep-lat]].
		
		
						\bibitem{DrachLattice21}
		V.~Drach, P.~Fritzsch, A.~Rago and F.~Romero-L\'opez,
		in \emph{38th International Symposium on Lattice Field Theory} (2021).
		
		
		\bibitem{Athenodorou:2014eua} 
		A.~Athenodorou, E.~Bennett, G.~Bergner and B.~Lucini,
		Phys.\ Rev.\ D {\bf 91}, no. 11, 114508 (2015)
		[arXiv:1412.5994 [hep-lat]].
		
		\bibitem{Bennett:2021ivn}
		E.~Bennett, A.~Athenodorou, G.~Bergner and B.~Lucini,
		[arXiv:2110.12979 [hep-lat]].
		
		\bibitem{Ryttov:2008xe} 
		T.~A.~Ryttov and F.~Sannino,
		Phys.\ Rev.\ D {\bf 78}, 115010 (2008)
		[arXiv:0809.0713 [hep-ph]].
		
		\bibitem{Kanazawa:2019tnf}
		T.~Kanazawa and M.~\"Unsal,
		Phys. Rev. D \textbf{102} (2020) no.3, 034013
		[arXiv:1909.05222 [hep-th]].
		
				
		\bibitem{Bergner:2014dua} 
		G.~Bergner and S.~Piemonte,
		JHEP {\bf 1412}, 133 (2014)
		[arXiv:1410.3668 [hep-lat]].
		
		
		
		\bibitem{Bergner:2018unx} 
		G.~Bergner, S.~Piemonte and M.~Ünsal,
		JHEP {\bf 1811}, 092 (2018)
		[arXiv:1806.10894 [hep-lat]].
		
		
		\bibitem{Bergner:2020mwl}
		G.~Bergner and S.~Piemonte,
		Phys. Rev. D \textbf{103} (2021) no.1, 014503
		[arXiv:2008.02855 [hep-lat]].
		
		\bibitem{Hasenfratz:2016gut} 
		A.~Hasenfratz, C.~Rebbi and O.~Witzel,
		Phys.\ Lett.\ B {\bf 773}, 86 (2017)
		[arXiv:1609.01401 [hep-ph]].

	\end{thebibliography}
\end{document}